\documentstyle[11pt,aaspp4]{article}
\slugcomment{Draft: ~~\today}

\begin{document}
\title{Multi-Read Out Data Simulator}
\author{J.\ D.\ Offenberg\altaffilmark{1}, 
R.\ J.\ Hanisch\altaffilmark{2},
D.\ J.\ Fixsen\altaffilmark{1}, 
H.\ S.\ Stockman\altaffilmark{2}, 
M.\ A.\ Nieto-Santisteban\altaffilmark{2}, 
R.\ Sengupta\altaffilmark{1},
J.\ C.\ Mather\altaffilmark{3}}

\altaffiltext{1}{Raytheon ITSS}
\altaffiltext{2}{Space Telescope Science Institute}
\altaffiltext{3}{NASA's Goddard Space Flight Center, Code 685}


\begin{abstract}
We present a data simulation package designed to create a series of
simulated data samples for a detector with non-destructive sampling
capability.  The original intent of this software was to provide a
method for generating simulated images for the Next Generation Space
Telescope, but it is general enough to apply to almost any
non-destructive detector or instrument.  MultiDataSim can be used to
generate ``practice'' sampling strategies for an instrument or
field-of-view, and thus can be used to identify optimal observation
strategies.
\end{abstract}

\section{Introduction}

MultiDataSim was written to test a cosmic--ray
identification--and--rejection program (Offenberg, et.\ al., 1999),
written for the Next Generation Space Telescope (NGST).  However, it
is useful for many purposes and for other observatories.  One use of
MultiDataSim is to compare on the merits of various data sampling
techniques and timing approaches when presented with a detector
capable of non-destructive reads.  By reading the detector many times,
it is possible to manipulate the data in ways that would otherwise be
unavailable, but defining an optimum strategy can become complicated.
A study of this trade-space (Fixsen, et.\ al., in press) suggests that
there is no general solution to this question: it depends on the
observing program and the specific parameters of the observation.

This paper is divided into three parts.  The first is a narrative
background description of MultiDataSim and its potential uses.  The
second section provides the documentation for the software.  The final
section, the appendix, contains the C++ source code for the software
itself.

These studies are supported by the NASA Remote Exploration and
Experimentation Project (REE), which is administered at the Jet
Propulsion Laboratory under Dr. Robert Ferraro, Project Manager.

\section{Background}

There are several methods for observing a single field-of-view when a
non-destructive capable-detector is used.  We discuss two of them here
(more information can be found in Fixsen, et.\ al., in press).  

The first is Fowler Sampling (Fowler \& Gatley, 1990), in which the
detector is sampled $N/2$ times at the start of the observation,
followed by a long integration and is then sampled $N/2$ times at the
end of the observation.  Essentially, we are measuring the
accumulating charge $N/2$ times, so we can reduce the random read
noise introduced by the detector by a factor of $\sqrt{N/4}$.

A second method is to sample the data Up-the-Ramp, at set intervals
during the observation---sampling at uniform intervals being the
simplest and most straightforward case.  The signal is then computed
as the slope of the ramp via a least-squares fit.  The read-noise
reduced by a factor of $\sqrt{N/12}$ (in the uniform-weighted case).
Up-the-Ramp sampling also allows observers to search the ramp for
glitches (e.g. those caused by cosmic rays), saturated pixels, drift
in the tracking, transient jitter and so on.  Such suspect data could
then be excluded from the processing while preserving the rest of the
data.  As this is just one possible trade, it is easy to see that
Fowler vs Up-the-Ramp vs some other sampling method can be a
complicated trade space.

The MultiDataSim tool allows the user to select an arbitrary set of
sampling times; Uniform and Fowler sampling algorithms are already
built into the code.  The software starts with a noiseless "Real Sky"
image and generates simulated observations for the specific intervals
with appropriate noise contamination.  Each sample is affected by
Poisson noise (i.e. photon--counting statistics), random read noise
and cosmic rays.  Most of the default values are chosen for the NGST.
Thus, the cosmic ray rate (5 particles~cm$^{-2}$~s$^{-1}$) is that
expected for a deep-space (L2) orbit, the physical pixel size is the
baseline for NGST ($27\mu \times 27\mu \times 10\mu$), and so on.  It
is possible to turn off these features, and override the default
values on the command-line.  The output images are generated as a
sequence of FITS (Flexible Image Transport System) files, one sample
of the detector array per file.

There are two functions which the authors have not yet had the
opportunity to implement in MultiDataSim, although work-arounds exist
for each.  The first is to apply a non-linearity curve to the data,
although this can be applied after the fact.  MultiDataSim also does
not apply a point-spread function (PSF) to the data.  This can easily
be done by applying the PSF to the input image.

\section{Documentation}

\noindent Calling sequence:\\
\indent MultiDataSim [options] {\it inputfile} [options]

where {\it inputfile} is a FITS file containing the ``ideal'' input
image.  Most of the options are order-independent and can appear
before or after the {\it inputfile} argument.

Options:

-co {\it x} \hfill\parbox[t]{5in}{Specifies a constant DC offset (versus dark current) to be applied to the image (default: 0.0).}

-craa \hfill\parbox[t]{5in}{Specify to use cosmic--ray area--of--effect.  This turns on a small mask so \~1\% of the cosmic ray flux ends up in each of the 4 adjacent--neighbor pixels and .1\% into the 4 diagonal-neighbor pixels.}

-crr {\it x} \hfill\parbox[t]{5in}{Specifies cosmic ray rate as {\it x}~events~cm$^{-2}$~sec$^{-1}$ (default: 4).}

-dc {\it x} \hfill\parbox[t]{5in}{Specifies the DC dark current is {\it x}~e$^{-}$~sec$^{-1}$ (default: 0.02).}

-df {\it file} \hfill\parbox[t]{5in}{Specifies the path to the image containing the dark frame (default: dark frame is blank).}

-expf {\it file} \hfill\parbox[t]{5in}{Specifies that the software should use a customized sample times instead of either Fowler or Uniform sampling.  {\it file} is the name of an ascii file containing a list of sample times (one per line).  The software will record a t=0 image as well as the ones specified.  Using the -expf keyword overrides the "-fs", "-n", and "-t" keywords.}

-ff file \hfill\parbox[t]{5in}{Specify Flat field image (default: none.)}

 -g {\it x} \hfill\parbox[t]{5in}{Specify the gain in e$^{-}$ data--unit$^{-1}$ (default: 4).}

-fs \hfill\parbox[t]{5in}{Specifies that the software should use Fowler Sampling instead of the default Uniform Sampling method.  The interval between Fowler samples is 5 seconds.}

-ie {\it x} \hfill\parbox[t]{5in}{Specifies the exposure time of the ``ideal'' input image in seconds (default: 1000).}

-n {\it x} \hfill\parbox[t]{5in}{Specifies that the observing sequence contains {\it x} observations including an observation at t=0 seconds (default: 64).}

-nc \hfill\parbox[t]{5in}{Specifies that the software should not inject cosmic rays into the detector.}

-ncf \hfill\parbox[t]{5in}{Specifies that the software should not record the injected cosmic ray events in separate files (see ``-o'').}

-np \hfill\parbox[t]{5in}{Specifies that the software should not apply Poisson noise to the data.}

-nr \hfill\parbox[t]{5in}{Specifies that the software should not apply readnoise to the data.}

-o {\it file} \hfill\parbox[t]{5in}{Specifies the base file name
for the output sequence (default: "datasim\_").  Each simulated data
sample is stored in its own FITS file named {\it fileXXXXXX.fit},
where {\it XXXXXX} is the time of the sample during the integration,
in whole seconds.  Note that this notation imposes 2 limits: elapsed
time between samples must be a minimum of 1 second, and the observing
sequence as a whole is limited to $10^6$ seconds.  In addition to
creating {\it fileXXXXXX.fit}, datasim also records cosmic ray events
in a series of files {\it fileXXXXXX.fit\_cosmic} (unless the -ncf
argument is specified).}

-pix {\it l,w,d} \hfill\parbox[t]{5in}{Specify length, width, depth of pixel in $\mu$ (default: 27,27,10).}

-r {\it x} \hfill\parbox[t]{5in}{Specify read noise in e$^{-}$ (default: 15).}

-sc {\it x} \hfill\parbox[t]{5in}{Specifies angular shield coverage is {\it x} degrees (default: 1.0).}

-t {\it x} \hfill \parbox[t]{5in}{Specifies that the observing sequence lasts {\it x} seconds (default: 1000).}

-v \hfill\parbox[t]{5in}{Specifies that the software should run in ``Verbose'' mode.}

\begin{appendix}
\begin{center}{\bf APPENDIX}\end{center}

The code here is long enough as it is, so we have omitted a few
functions and files from this appendix.  We include the header files
which define C++ classes, but omit all other header files.  The other
header files contain just procedure prototype lines, so recreating
them is straightforward.

The {\it ran1} function (called in cosmic.cc) and the {\it poidev} and
{\it gasdev} functions (called in detectorimage.cc) are all from the
Numerical Recipes library, and can be found in the text {\bf Numerical
Recipes} (Press, Flannery, Teukolsky \& Vetterling).  We are using a
version of the on-line library based on the 1986 edition of {\bf
Numerical Recipes}; unfortunately, that version of the library is
Fortran-based, so the translation to C was awkward in places.  We
created a C library, librecipes\_g++, containing {\it ran1, poidev}
and {\it gasdev}, plus the Numerical Recipes functions upon which
these 3 depend.  In addition to the Numerical Recipes library, we are
also using the standard C math library.

Implementing FITS I/O is left as an exercise for the reader.  The {\it
-lcfitsio} in the Makefile refers to the ``CFITSIO'' library.  The
CFITSIO library is available from other sources on the web and is not
included in this paper.  We are using version 1.4, by William Pence (1998).

\section{Makefile}

\begin{verbatim}
# MAKEFILE for MultiDataSim
#
#	Makefile to generate the MultiDataSim software.  This Makefile was
#	written for the standard Sun Solaris(tm) Make utility.
#
# Written by Joel D. Offenberg, Raytheon ITSS
#
OBJ 	= datasim.o detectorimage.o fits.o string2num.o cosmic.o 
INC_DIR	= -I../include 
LIB_DIR = -L../lib -L/usr/local/lib
LIBS = -lcfitsio -lm -lrecipes_g++
FLAGS = -O
CCC	= g++  $(FLAGS)

MultiDataSim : $(OBJ) 
	$(CCC) -o MultiDataSim $(OBJ) $(INC_DIR) $(LIB_DIR) $(LIBS)

install : MultiDataSim
	mv MultiDataSim ../bin
	chmod 755 ../bin/MultiDataSim

.cc.o:
	$(CCC) -c -o $*.o $*.cc $(INC_DIR) 

#
\end{verbatim}

\section{pixel.h}
\begin{verbatim}
/*****

      PIXEL.H

      This is a simple class to encapsulate several properties of a pixel in
the detector.  At this writing, it is just the size, although we might 
end up encapsulating other properties over time for neatness' sake.

      Written by Joel Offenberg, Raytheon ITSS

******/

#ifndef PIXEL_H
#define PIXEL_H

class PIXEL 
{
 public:
  float x, y, z;
  PIXEL() { x = 27.; y=27.; z=10.; }
  PIXEL(float a,float b,float c) { x = a; y = b; c = z; }
};

#endif

\end{verbatim}

\section{detectorimage.h}
\begin{verbatim}
/******
  detectorimage.h

  File containing the class definitions for the DetectorImage class.

  Written by Joel D. Offenberg, Raytheon ITSS

  *****/

#ifndef DETECTOR_IMAGE_H
#define DETECTOR_IMAGE_H

#include "pixel.h"

class DetectorImage 
{
private:
  float **image;
  long axis1;
  long axis2;
  float exptime;
  float deltatime;
public:
  DetectorImage(int, int);  
  DetectorImage(float *, int, int, float);
  DetectorImage(DetectorImage* );
  DetectorImage(char *filename);
  DetectorImage(char *filename, long startrow, long startcol, 
		long numrow, long numcol);
  ~DetectorImage();

  void zero_out();
  
  void setval(int i, int j, float value) { image[i][j] = value; }
  float getval(int i, int j) { return image[i][j]; }

  float *Buffer();
  long Axis1() { return axis1; }
  long Axis2() { return axis2; }
  float exposure() { return exptime; }
  void set_exposure(float exp) { exptime = exp; }
  float get_deltatime() {return deltatime;}
  void scale_to_exptime(float t);
  void counting_statistics();
  void readnoise(float);
  void roundoff(float);
  void roundoff() { this->roundoff(1.); }

  void write(char *filename);
  void write_int(char *filename);
  void plus_equals(DetectorImage);
  void plus_equals(DetectorImage *);
  void plus_equals(float);
  void divide_equals(float c);
  void times_equals(float c);
  void times_equals(DetectorImage *A);
  void times_equals(DetectorImage A) {times_equals(&A);};
  void ceiling(float c);

  void add_cosmicrays() 
		{ add_cosmicrays(exposure()); }
  void add_cosmicrays(float time) 
		{add_cosmicrays(time, 1.0);}
  void add_cosmicrays(float time, float shieldcover) 
		{add_cosmicrays(time,shieldcover,4.);}
  void add_cosmicrays(float time, float shieldcover, float crrate) 
		{add_cosmicrays(time,shieldcover,crrate,0);}
  void add_cosmicrays(float time, float shieldcover, float crrate, char mask) 
		{add_cosmicrays(time, shieldcover,crrate,mask,*(new PIXEL()));}
  void add_cosmicrays(float time, float shieldcover, float crrate, 
							char mask, PIXEL P);

  void test_for_infinity(char *message);
};

#endif

\end{verbatim}

\section{detectorimage.cc}
\begin{verbatim}
/*****
  detectorimage.c

  C code files for the DetectorImage class.  See detectorimage.h.

  Written by Joel D. Offenberg, Raytheon ITSS
  *****/


#include <stdio.h>
#include <string.h>
#include <math.h>
#include <nr.h>
#include "fitsio.h"
#include "pixel.h"
#include "string2num.h"
#include "fits.h"
#include "detectorimage.h"
#include "cosmic.h"

extern long idnum;

/** Constructors **/

/*  Create an empty image naxis1 x naxis2 */
DetectorImage::DetectorImage(int naxis1, int naxis2) {
  int j,i;

  image = new (float*)[naxis1];
  for (j=0; j<naxis1; j++) {
    image[j] = new float[naxis2];
    for (i=0; i<naxis2; image[j][i++] = 0.0);
  }

  axis1 = naxis1;
  axis2 = naxis2;
  exptime = 1000.;
}

/*  Create an image and fill with naxis1 x  naxis2 elements from ``buffer.'' */
DetectorImage::DetectorImage(float *buffer, int naxis1, int naxis2, float exp) {
  int i,j;
  long count=0;

  image = new (float*)[naxis1];
  for (j=0; j<naxis1; j++) {
    image[j] = new float[naxis2];
    for (i=0; i<naxis2; image[j][i++] = buffer[count++]);
  }
    
  axis1 = naxis1;
  axis2 = naxis2;
  exptime = exp;
}

/*  Make a duplicate image */
DetectorImage::DetectorImage(DetectorImage *A) {
  float *buffer;
  int i,j;
  long count=0;

  image = new (float*)[A->Axis1()];
  for (j=0; j<A->Axis1(); j++) {
    image[j] = new float[A->Axis2()];
    for (i=0; i<A->Axis2(); i++) {
      image[j][i] = A->image[j][i];
    }
  }
    
  axis1 = A->Axis1();
  axis2 = A->Axis2();
  exptime = A->exposure();
}  

/*  Read image from a FITS file on the disk */
DetectorImage::DetectorImage(char *filename) {
  float *buffer;
  long naxis[2], count=0;
  int i,j;
  char *saveexp;

  if ( (saveexp = readheader(filename,"EXPOSURE")) == NULL) {
    if ((saveexp = readheader(filename,"EXPTIME")) == NULL) {
      exptime = 1.0;
    } else {
      saveexp = strchr(saveexp,'=');
      exptime = string2num(saveexp);
    } 
  } else {
    saveexp = strchr(saveexp,'=');
    exptime = string2num(saveexp);
  }

  if ((saveexp = readheader(filename,"DELTATIM")) != NULL) {
    saveexp = strchr(saveexp,'=');
    deltatime = string2num(saveexp);
  } else deltatime = 1.0;
  buffer = readimage(filename, naxis);
  image = new (float*)[naxis[0]];
  for (j=0; j<naxis[0]; j++) {
    image[j] = new float[naxis[1]];
    for (i=0; i<naxis[1]; image[j][i++] = buffer[count++]);
  }
    
  axis1 = naxis[0];
  axis2 = naxis[1];

  delete buffer;
}

/*  Read a section from a FITS file on the disk. */
DetectorImage::DetectorImage(char *filename, long startrow, long startcol, 
						long numrow, long numcol) {
  float *buffer;
  long naxis[2], count=0;
  int i,j;
  char *saveexp;

  if ( (saveexp = readheader(filename,"EXPOSURE")) == NULL) {
    if ((saveexp = readheader(filename,"EXPTIME")) == NULL) {
      exptime = 1.0;
    } else {
      saveexp = strchr(saveexp,'=');
      exptime = string2num(saveexp);
    } 
  } else {
    saveexp = strchr(saveexp,'=');
    exptime = string2num(saveexp);
  }

  if ((saveexp = readheader(filename,"DELTATIM")) != NULL) {
    saveexp = strchr(saveexp,'=');
    deltatime = string2num(saveexp);
  } else deltatime = 1.0;
  buffer = readsubimage(filename, naxis, startrow, startcol, numrow, numcol);
  image = new (float*)[naxis[0]];
  for (j=0; j<naxis[0]; j++) {
    image[j] = new float[naxis[1]];
    for (i=0; i<naxis[1]; image[j][i++] = buffer[count++]);
  }
    
  axis1 = naxis[0];
  axis2 = naxis[1];

  delete buffer;
}
    

/*** Destructor ****/
DetectorImage::~DetectorImage() {
  int i;

  for (i=0; i<this->Axis1(); i++) {
    delete image[i];
  }

  delete image;
};



/*** Buffer ***/
/* Returns a pointer to a single buffer which contains the entire image in
   one long stretch instead of the standard 2-D array that the ``image'' 
   contains.
*/

float* DetectorImage::Buffer() {
  float *result;
  int j;

  result = new float[axis1*axis2];

  for (j=0; j<axis1; j++) { 
    memcpy( (result+j*axis2), image[j], axis2*sizeof(float));
  }

  return result;
}


/*** plus_equals ***/
/*
  Performs the += operation on two DetectorImages or a detector image 
  plus an offset.  I did it this way (instead of overloading the += operator)
  because it removed an ambiguity when dealing with a pointer to a 
  DetectorImage.
*/

/*  Plus_equals with an image ``a''. */
void DetectorImage::plus_equals(DetectorImage a) {
 
  int i,j;

  if ((a.Axis1() != this->Axis1()) || (a.Axis2() != this->Axis2())) {
    printf("\n These images are not the same size!");
    return;
  }
  
  for (i=0; i<this->Axis1(); i++) {
    for (j=0; j<this->Axis2(); j++) {
      this->image[i][j] += a.image[i][j];
    }
  }
  exptime += a.exposure();
}

/*  Plus_equals with a reference to an image ``*a''.  */
void DetectorImage::plus_equals(DetectorImage *a) {
 
  int i,j;


  if ((a->Axis1() != this->Axis1()) || (a->Axis2() != this->Axis2())) {
    printf("\n These images are not the same size!");
    return;
  }
  

  for (i=0; i<this->Axis1(); i++) {
    for (j=0; j<this->Axis2(); j++) {
      this->image[i][j] = this->image[i][j] + a->image[i][j];
    }
  }
  exptime += a->exposure();
}

/*  Plus_equals with a single value. */
void DetectorImage::plus_equals(float offset)
{
  int i,j;

  for (i=0; i<this->Axis1(); i++) {
    for (j=0; j<this->Axis2(); j++) {
      this->image[i][j] += offset;
    }
  }
}

/**** DetectorImage::roundoff ****/
/* 
Rounds off image data to the nearest n-th integer.  Also cuts off the maximum
value at 2^16, to prevent 16-bit integer overflow.
 */

void DetectorImage::roundoff(float n) {
  
  int i,j;

  for (i=0; i<this->Axis1(); i++) {
    for (j=0; j<this->Axis2(); j++) {
      image[i][j] = (image[i][j]>65535.)?65535. :(float)((int) image[i][j]/n)*n;
    }
  }
}

/**** DetectorImage::write ****/
/* Saves the image to disk as a FITS file. */

void DetectorImage::write(char *filename) {
  long naxis[2];
  float *buffer;
  
  buffer = this->Buffer();
  naxis[0] = this->Axis1();
  naxis[1] = this->Axis2();
  
  writeimage(buffer, this->exposure(), filename, naxis);
  delete buffer;
  
  return;
}

/**** DetectorImage::write_int ****/
/* Saves the image to disk as an INTEGER FITS file. */

void DetectorImage::write_int(char *filename) {
  long naxis[2];
  float *buffer;
  
  buffer = this->Buffer();
  naxis[0] = this->Axis1();
  naxis[1] = this->Axis2();
  
  writeimage_int(buffer, this->exposure(), filename, naxis);
  delete buffer;
  
  return;
}

/**** divide_equals ****/
/* A routine to divide the image (and exposure time) by a scalar constant.  
	It was done this way (instead of overloading /=) to avoid ambiguity 
	with pointer arithmetic. 
*/
void DetectorImage::divide_equals(float c) {
  int i,j;

  for (i=0; i<Axis1(); i++) {
    for (j=0; j<Axis2(); j++) {
      image[i][j] /= c;
    }
  }
  
  exptime /= c;
}

/**** times_equals ****/
/* A routine to multiply the image (and exposure time) by a scalar constant.  
	It was done this way (instead of overloading *=) to avoid ambiguity 
	with pointer arithmetic. 
*/
void DetectorImage::times_equals(float c) {
  int i,j;

  for (i=0; i<Axis1(); i++) {
    for (j=0; j<Axis2(); j++) {
      image[i][j] *= c;
    }
  }
  
  exptime *= c;
}

/**** times_equals ****/
/* A routine to multiply the image by another image. 
	It was done this way (instead of overloading *=) to avoid ambiguity 
	with pointer arithmetic. 
*/

void DetectorImage::times_equals(DetectorImage *A) {
  int i,j;

  for (i=0; i<Axis1(); i++) {
    for (j=0; j<Axis2(); j++) {
      image[i][j] *= A->getval(i,j);
    }
  }
}


/**** scale_to_exptime ****/
/* A routine to scale the image and exposure time to the specified number
   of seconds of exposure time.  Basically, this uses divide_equals to 
   do the division, but it calculates the factor based on the existing 
   exposure time rather than making the user do it.

   */

void DetectorImage::scale_to_exptime(float t) {
  
  t = this->exposure() / t;
  
  this->divide_equals(t);
}

DetectorImage *emptyimage;

/**** add_cosmicrays(float time, float shieldcover, float rate) ****/
/* Routine to add cosmic rays to the image.  Asks the user to supply the
   number of seconds of integration for the cosmic rays.  Ultimately,
   the proper routine will be add_cosmicrays(), which will extract the 
   current exposure time and use that to calculate the number of CRs.
   */

void DetectorImage::add_cosmicrays(float time, float shieldcover, 
				float crrate, char mask, PIXEL P) {
  
  float **crimage;
  int i,j;
  float tmp;

  /* Initialize CRImage */
  crimage = new (float *)[this->Axis1()];
  for (j=0; j<this->Axis1(); j++) {
    crimage[j] = new (float)[this->Axis2()];
    for (i=0; i<this->Axis2(); i++) {
      crimage[j][i] = 0.0;
    }
  }

  
  /* Compute cosmic rays.  We'll need to tweak some of the instrumental
     parameters for cosmic(); see cosmic.c.
     */

  cosmic(crimage,this->Axis1(),this->Axis2(), time, shieldcover,crrate,mask, P);

  for (i=0; i<this->Axis1(); i++) {
    for (j=0; j<this->Axis2(); j++) {
      image[i][j] += crimage[i][j];
      emptyimage->setval(i,j, emptyimage->getval(i,j)+crimage[i][j]);
    }
  }
  
  for (j=0; j<this->Axis1(); j++) {
    delete crimage[j];
  }
  delete crimage;
}


/*** counting_statistics ***/
/* Takes "ideal" image and generates a version based on Poisson counting 
   statistics. */

void DetectorImage::counting_statistics() {
  int i,j;
  
  for (i=0; i<Axis1(); i++) {
    for (j=0; j<Axis2(); j++) {
      image[i][j] =  poidev(image[i][j], &idnum);
    }
  }
}

/*** readnoise ***/
/* Takes "ideal" image and generates a version based on Gaussian distribution
   statistics. */

void DetectorImage::readnoise(float variance) {
  int i,j;
  float tmp;

  for (i=0; i<Axis1(); i++) {
    for (j=0; j<Axis2(); j++) {
      image[i][j] += (float) ((int)gasdev(&idnum) * variance);
    }
  }
}

/*** zero_out ***/
/* Simple routine to erase an image.  Also sets exptime to zero. */

void DetectorImage::zero_out() {
  int i;
  int j;

  for (i=0; i<Axis1(); i++) {
    for (j=0; j<Axis2(); j++) {
      image[i][j] = 0.;
    }
  }
  
  exptime = 0.0;
}

/*
  DetectorImage:ceiling

  Limit value at a top value (i.e. saturation limit).
*/

void DetectorImage::ceiling(float c) {
  int i,j;
  
  for (i=0; i<Axis1(); i++) {
    for (j=0; j<Axis2(); j++) {
      if (image[i][j] >= c) {
	image[i][j] = c;
      }
    }
  }
}


/*** test_for_infinity ***/ 
/* 
	A routine to test if the image contains any ``infinite'' (or
	impossibly large) values.  Debugging routine.
*/
void DetectorImage::test_for_infinity(char *message) {
  int i,j;

  for (i=0; i<Axis1(); i++) {
    for (j=0; j<Axis2(); j++) {
      if (fabs(image[i][j] > 1e+20)) {
	printf("%s \n",message);

	//Short-circuit out of loop if one is found
	i = Axis1();
	j = Axis1();
      }
      
    }
  }
}

\end{verbatim}

\section{cosmic.cc}
\begin{verbatim}
/**********
   cosmic.cc

   C Code file containing routines to generate simulated cosmic ray hits
   on detector.  The entry point for these routines is cosmic:

   cosmic(detector_array, array_xsize, array_ysize, length)
       
   detector_array[array_ysize][array_xsize] is a 2-dimensional array to which
            the cosmic rays will be added.
   array_xsize, array_ysize are the dimensions of the detector_array
   length is the number of seconds over which cosmic rays are to be added.

   ASSUMPTIONS:
   1.  All CR's liberate 100 (+/- 10) electrons per 0.1 micron travel.  10% 
       of the cosmic rays are He nuclei, and thus have 4x the effect.

   2.  The plane of the detector (+/- some extra) is blocked by a shield.

   3.  The CR sheild stops 100% of CRs, regardless of energy.

   4.  A CR event in the detector has no lasting effect.

  Written by: Joel D. Offenberg, Raytheon ITSS

   **********/


#include <sys/types.h>
#include <time.h>
#include <math.h>
#include <stdlib.h>
#include <nr.h>
#include "pixel.h"
#include "cosmic.h"

extern long idnum;

/***
  Global constants
***/
const float deg2rad = 3.14159265/ 180.;  /* Number of radians per degree */

/*****
  cosmic
  
  Routine to deliver simulated Cosmic ray hits on the detector.  
 
  INPUTS:
  int **image:  2-dimensional array containing image to which cosmic rays
                are to be added.  
  int xs, ys:   Size of the image, in pixels, to which the cosmic rays are to
                be added.
  float time:   Number of seconds of integration for the image.
  float shieldcover    Cutoff angle (in degrees) due to shield
                       assume shield is perfect 

  OUTPUTS:
  The image array is updated.
  Function returns a "1" upon success.

  *****/

extern int VERBOSE;

int cosmic(float **image, int xs, int ys, float time, float shieldcover, 
	float rate, char mask, PIXEL P) {

  float edl = 100.;        /* # electrons liberated per micron travel of H */

  float n_rays;              /* # cosmic rays on detector */
  int i,j,k,m,p,q;                     
  float xpos, ypos, zpos;     /* position of cosmic ray hit on detector */
  float theta, cosphi,sinphi; /* direction angles (and cosine thereof) of 
				 cosmic ray velocity */
  float pathlength;          /* Total path length through detector */
  int npath;                  /* number of micron steps needed to pass 
				 through detector */
  float dpath;               /* Number of microns per step.  In theory,
				 this is 1 by definition, except for the last
				 step, but instead we seem to be assuming 
				 even not-quite-micron steps.  */
  float depath;              /* Number of electrons liberated per step. */
  float dx, dy, dz;          /* Amount of motion per step */
  float charge;  
  int NHe = 0;

  n_rays = xs * P.x * 1e-4 * ys * P.y * 1e-4 * rate * time;

  n_rays = poidev(n_rays, &idnum);    /* Add poisson statistics.  */
  if (n_rays < 0) { n_rays = 0; }   /* Correct for negative number */

  for (i=0; i<n_rays; i++) {

    /* Generate random numbers between 0..xs-1 and 0..ys-1. */
    xpos = ran1(&idnum) * (float) xs;
    ypos = ran1(&idnum)* (float) ys;
    zpos = 0.0;
    
    theta = ran1(&idnum)*2*3.141592653;

    sinphi = sqrt(ran1(&idnum));
    cosphi = sqrt(1.0 - sinphi*sinphi);

    if (asin(sinphi)/deg2rad > shieldcover) {

    /* 10% of the time, the CR will be a He nucleus, with 4 times the effect */
    charge = ( ran1(&idnum) > 0.9) ? 2. : 1.;  
    if (charge == 2.) {NHe++;}

    depath = 100.;
    dpath = 0.1;

    dx = cos(theta) * dpath * sinphi;    /* x element per step */
    dy = sin(theta) * dpath * sinphi;    /* y element per step */
    dz = dpath * cosphi;                 /* z element per step */

    /* Compute position of cosmic ray at each step and add appropriate 
       signal to detector array.  Stop when outside the detector. */

    for (zpos = 0.0; (zpos <= P.z) && (xpos>=0) && (ypos>=0) && 
					(xpos<xs) && (ypos<ys); zpos += dz) {
      if (mask == 1) {
	// Add cosmic ray area-of-effect mask.  The values are based on 
	// Bernie Rauscher's cosmic ray measurements.  
	int xxpp,yypp;
	float crmask[3][3] = {{1.06e-3,1.66e-2,1.06e-3},
			      {1.66e-2,1.00,1.66e-2},
			      {1.06e-3,1.66e-2,1.06e-3}};
	float crcont = depath*(charge*charge);

	for (xxpp=0; xxpp<=2; xxpp++) {
	  for (yypp=0; yypp<=2; yypp++) {
	    if ((xxpp+xpos-1 >= 0) && (yypp + ypos-1 >= 0) && 
		(xxpp+xpos-1 < xs) && (yypp + ypos-1 < ys)) {
	      image[(int) xpos + xxpp-1][(int) ypos+yypp-1] += 
			poidev(crcont, &idnum) * crmask[xxpp][yypp];
	} } } } 
      else 
	image[(int) xpos][(int) ypos] += poidev(depath*(charge*charge), &idnum);
      xpos += dx/P.x;
      ypos += dy/P.y;
    }

    /* If the CR hits the bottom of the detector the odds are it won't hit it
       on an even increment in the steps above.  Thus, add an extra term for 
       the last little bit.  The CR-area PSF is not applied here, since this
       is just a little bit at the end.  [It probably should be, but we'll
       live for now.]
    */
    if ((xpos>=0) && (ypos>=0) && (xpos<xs) && (ypos<ys)) {
      image[(int) xpos][(int) ypos] += 
	poidev(depath*(charge*charge)*(P.z-zpos+dz)/dz, &idnum);
    }
    }
  }
  

  
  if (VERBOSE) printf("\t%d CRs, %d He\n",(int) n_rays, NHe);

  /* DONE */
  return 1;
  
}


\end{verbatim}

\section{datasim.cc}
\begin{verbatim}

/*****
  Main.c

  Main file for data NGST warts-and-all data simulator.

  Written by: Joel D. Offenberg, Raytheon ITSS

  *****/

#include <string.h>
#include <sys/types.h>
#include <time.h>
#include <stdio.h>
#include "pixel.h"
#include "fits.h"
#include "detectorimage.h"
#include "string2num.h"
#include "datasim.h"

long idnum;

int VERBOSE=0;

extern DetectorImage *emptyimage;

/*** main ***/
/*
	the entry point and controlling routine.
*/

int main(int argc, char **argv) {
  
  char filename[80];    //input file name

  // Image objects
  DetectorImage *IdealImage, *DeltaImage, *IntegrateImage, *IntegratePrime, 
    *FlatImage;   

  //Other file names.
  char outfilename[80], outfilebase[70], darkfile[80],flatfile[80];  
  float exptime;                    //exposure time in seconds.
  float numimg;                      //number of intervals.
  float duration;                   //keep track of exposure time to date.
  float *deltatimes;                //List of exposure-time intervals.
  float offset = 0.0;               //DC offset 
  float dccurrent=0.02;             //Dark current (e-/s)
  float initexp = 1000;             //Exposure-time of original image.
  float shieldcover = 1.0;          //Angular coverage of shield around plane.
  float crrate = 4.;                //Number of cosmic rays/s/cm/cm.
  float rnvar = 15.;                //Readnoise.
  float gain = 4.;                  //gain (e-/data unit)
  char cosmicbool=0, readnoise=0, cosmicout=0, exptimefile[100];
  char tempstring[80];
  int count;
  FILE *inputfile;
  PIXEL *P;

  P = new PIXEL();

  if (parse_args(argc, argv, filename, outfilebase, darkfile, flatfile, 
	&offset, &duration, &numimg, &cosmicbool, &readnoise, &cosmicout, 
	&dccurrent, &initexp, &shieldcover, exptimefile, &crrate, &rnvar, 
	&gain, P)) {
    return 0;
  }

    /* Read in source image.  Assume that exposure time is 1000 seconds.  */
  if (strcmp(filename,"") == 0) {
    IdealImage = new DetectorImage(1024,1024);
  } else {
    IdealImage = new DetectorImage(filename);
    IdealImage->set_exposure(initexp);
  }

  /* Read in flatfield image.  If not specified, make it a constant image. */
  if (strcmp(flatfile,"") == 0) {
    FlatImage = new DetectorImage(IdealImage->Axis1(),IdealImage->Axis2());
    FlatImage->plus_equals(1.);
  } else {
    FlatImage = new DetectorImage(flatfile);
  }

  emptyimage = new DetectorImage(IdealImage->Axis1(), IdealImage->Axis2());

  //Seed the random number generator: use the system clock.
  idnum = -time(0) ;

  /* Initialize and zero-out IntegrateImage.  This is a little wasteful
     from a computation standpoint, but it guarantees that IntegrateImage
     will be the right size. 

     Modified to include possibility that a non-zero dark and/or a constant
     offset should be applied to the image.  */

  if (strcmp(darkfile,"") == 0) {
    IntegrateImage = new DetectorImage(IdealImage);
    IntegrateImage->zero_out();
  } else {
    IntegrateImage = new DetectorImage(darkfile);
  }
  IntegrateImage->plus_equals(offset);

  //Write out darkfield image.  

  IntegratePrime = new DetectorImage(IntegrateImage);

  IntegratePrime->times_equals(FlatImage);

  if (readnoise == 0) { IntegratePrime->readnoise(rnvar); }

  //Modify for gain
  IntegratePrime->divide_equals(gain);
  IntegratePrime->set_exposure(0.0);
  IntegratePrime->roundoff();

  sprintf(outfilename,"%s%06d.fit",outfilebase,(int) 0);
  IntegratePrime->write_int(outfilename);

  //Subtract this image from the number of images, "numimg."
  numimg -= 1;

  if (cosmicout == 0) {
    sprintf(outfilename,"%s%06d.fit_cosmic",outfilebase, (int) 0);
    emptyimage->write_int(outfilename);
  }

  /* Figure out deltatimes.  If "-fs" (Fowler Sample) option was selected,
     first numimg/2 exposures are 5 second each, followed by 1 big exposure,
     then numimg/2 5 second exposures.  Otherwise, the delta times are 
     all evenly spaced. */

  deltatimes = new (float)[numimg];
  if ((cosmicbool & 8) != 0) {
    delete deltatimes;
    deltatimes = new (float)[1000];

    inputfile = fopen(exptimefile,"r");
    for (count=0; !feof(inputfile); count++) {
      fscanf(inputfile,"%s\n",tempstring);
      deltatimes[count] = string2num(tempstring);
    }
    numimg = count;
    duration = deltatimes[count];
    for (count=(int)numimg; count > 0; 
			deltatimes[count--] -= deltatimes[count-1]);
    fclose(inputfile);
  } else
  { if ((cosmicbool & 2) != 0) 
    {
      for (count = 0; count < (numimg)/2; deltatimes[count++] = 5.0);
      deltatimes[count++] = duration - 5.0*(numimg-1);
      for ( ; count < numimg; deltatimes[count++] = 5.0);
    } 
    else 
    {
      for (count=0; count<numimg; deltatimes[count++] = duration/numimg);
    }
  }


  /*** Repeat loop for equally-spaced observations through duration. ***/

  for (count=0,exptime=deltatimes[count]; count<numimg; 
				exptime+= deltatimes[++count]) {

    if (VERBOSE) printf("%d seconds\n",(int) exptime);

    /* Calculate ideal image scaled to exposure exptime. */
    DeltaImage = new DetectorImage(IdealImage);
    DeltaImage->scale_to_exptime(deltatimes[count]);
    DeltaImage->plus_equals(dccurrent*deltatimes[count]);

    /* Modify image by Poisson counting statistics as detected */
    if ((cosmicbool & 4) != 4) {
      DeltaImage->counting_statistics();
    }

    /* Add cosmic rays */
    if ((cosmicbool & 1) != 1) {
      emptyimage->zero_out();

      DeltaImage->add_cosmicrays(deltatimes[count],shieldcover,crrate,
					((cosmicbool & 16) != 0)?1:0,*P);
 
      sprintf(tempstring,"%06d",(int)exptime);
      DeltaImage->test_for_infinity(tempstring);
    } 
    DeltaImage->times_equals(FlatImage);
    /* Remember the history of the observation until now */
    IntegrateImage->plus_equals(DeltaImage);

    /* Copy full observation, add read noise (gaussian distribution,
       variance of 15 electron-units).
    */
    IntegratePrime = new DetectorImage(IntegrateImage);

    if (readnoise == 0) {
      IntegratePrime->readnoise(rnvar);
    }
    
    /* Generate filename and save the the observation as it has been
       simulated until now. */
    
    IntegratePrime->divide_equals(gain);

    IntegratePrime->set_exposure( IntegratePrime->exposure() *gain); 
    //This is needed because the previous step messes up the exptime.

    IntegratePrime->roundoff();

    sprintf(outfilename,"%s%06d.fit",outfilebase,(int) exptime);
    IntegratePrime->write_int(outfilename);

    if (cosmicout == 0) {
      sprintf(outfilename,"%s%06d.fit_cosmic",outfilebase, (int) exptime);
      emptyimage->write_int(outfilename);
    }

    /* Free up allocated space before next loop. */
    delete DeltaImage;
    delete IntegratePrime;
  }

  return 1;    //Done.
}

/*** parse_args ***/
/*
	Collect all of the argument-parsing routines into one place.  
	There are C-library routines that will do most of this, but,
	unfortunately, they don't exist on one of the platforms we
	needed to support.
*/

char parse_args(int argc, char **argv, char *inputfile, char *outputbase,
		char *darkfile, char *flatfile, float *offset, float *exptime,
		float *numimg, char *cosmicbool, char *readnoise, 
		char *cosmicout, float *dccurrent, float *initexp, 
		float *shieldcover, char *exptimefile, float *crrate, 
		float *rnvar, float *gain, PIXEL *P) {
  int i;
  FILE *input;

  /*  Intialize various outputs with default values. */

  *exptime = 1000.;     //Exp exptime = 1000s
  *numimg = 64.;      //64 readings within the obs time, numbered 0...63.
  strcpy(outputbase, "datasim_");    //Output files named datasim_xxxxxx.fit,
  // where xxxxxx is the integral number of seconds into the observation
  // which the data was taken.

  strcpy(inputfile, "");   //Default is an empty image (No file).
  strcpy(darkfile, "");    //Default dark field is an image
  
  for (i=1; i<argc; i++) {
    if (strcmp(argv[i], "-h") == 0) {
      printf("%s inputfile [-t exptime -n numobs -o outputbase -h]\n", 
	     argv[0]);
      return 1;
    } else 
    if (strcmp(argv[i], "-t") == 0) {
      *exptime = string2num(argv[i+1]);
      i+= 1;
    } else 
    if (strcmp(argv[i], "-n") == 0) {
      *numimg = string2num(argv[i+1]);
      i+=1;
    } else
    if (strcmp(argv[i], "-o") == 0) {
      strcpy(outputbase, argv[i+1]);
      i+=1;
    } else 
    if (strcmp(argv[i], "-df") == 0) {
      strcpy(darkfile,argv[i+1]);
      i+=1;
    } else
    if (strcmp(argv[i], "-ff") == 0) {
      strcpy(flatfile,argv[i+1]);
      i+=1;
    } else
    if (strcmp(argv[i], "-dc") == 0) {
      *dccurrent = string2num(argv[i+1]);
      i+=1;
    } else
    if (strcmp(argv[i], "-ie") == 0) {
      *initexp = string2num(argv[i+1]);
      i+=1;
    } else
    if (strcmp(argv[i], "-sc") == 0) {
      *shieldcover = string2num(argv[i+1]);
      i+=1;
    } else
    if (strcmp(argv[i], "-co") == 0) {
      *offset = string2num(argv[i+1]);
    } else
      if (strcmp(argv[i], "-v") == 0) {
	VERBOSE = 1;
    } else
    if (strcmp(argv[i],"-nc") == 0) {
      *cosmicbool += 1;
      printf("NO CRs\n");
    } 
    else 
    if (strcmp(argv[i],"-ncf") == 0) {
      *cosmicout = 1;
    }
    else
    if (strcmp(argv[i],"-fs") == 0) {
      *cosmicbool += 2;
      printf("Fowler Sampling\n");
    }
    else
    if (strcmp(argv[i],"-expf") == 0) {
      *cosmicbool += 8;
      strcpy(exptimefile, argv[++i]);
      printf("Custom Pattern: %s\n",argv[i]);
    }  
    else
    if (strcmp(argv[i],"-np") == 0) {
      *cosmicbool += 4;
      printf("No Poisson Noise\n");
    } 
    else
    if (strcmp(argv[i],"-craa") == 0) {
      *cosmicbool += 16;
      printf("Using Cosmic Ray current-leak.\n");
    }
    else
    if (strcmp(argv[i],"-nr") == 0) {
      *readnoise = 1;
      printf("No read noise\n");
    }
    else
    if (strcmp(argv[i],"-crr") == 0) {
      *crrate = string2num(argv[++i]);
    }
    else
    if (strcmp(argv[i],"-r") == 0) {
      *rnvar = string2num(argv[++i]);
    } else
    if (strcmp(argv[i],"-g") == 0) {
      *gain = string2num(argv[++i]);
    } else
    if (strcmp(argv[i],"-pix") == 0) {
      char a[100],*b=argv[++i];
      
      strncpy(a,b,strcspn(b,","));
      P->x = string2num(a);

      b += strcspn(b,",") +1;;
      strncpy(a,b,strcspn(b,","));
      a[strcspn(b,",")] = 0;
      P->y = string2num(a);

      b += strcspn(b,",")+1;
      P->z = string2num(b);

    }
    else
      strcpy(inputfile, argv[i]);
  }
  return 0;
}

\end{verbatim}

\section{string2num.cc}
\begin{verbatim}
/*******
  string2num.c
  
  C Code file containing a routine to convert a string to a number (floating
  point).


  Written by Joel D. Offenberg, Raytheon ITSS
  ******/

#include <string.h>
#include <math.h>

float string2num (char *string)
{
  float res=0.0, place=1.;
  int i, l;
  char dp = 0;

  l = strlen(string);
  for (i=0; i<l; i++) {
    switch (string[i]) {
    case '0':
    case '1':
    case '2':
    case '3':
    case '4':
    case '5':
    case '6':
    case '7':
    case '8':
    case '9':   
      if (dp == 1) {
	place /= 10;
	res += (string[i] - '0') * place;
      }
      else {
	res *= 10.0;
	res += (float) string[i] - '0';
      }
      break;
    case '.':
      dp = 1;
      break;
    case 'e':
    case 'E':
      if (string[i+1] == '-') {
	res /= pow(10.,string2num(string+i+2));
      } else {
	float tmp = 0, tmp2;
	tmp =  string2num(string+i+2);
	tmp2 = pow(10.0,tmp);
	
	res *= pow(10.,string2num(string+i+2));
      }
      i=l;
      break;
    default:
      break;
    }
  }				
  return res;
}


\end{verbatim}

\end{appendix}

\end{document}